\documentclass[oldversion]{aa}
\usepackage{amsmath}
\usepackage{txfonts}
\usepackage{graphicx}
\usepackage{natbib}
\bibpunct{(}{)}{;}{a}{}{,}
\newcommand{\ls}{LS IV-14$^\circ$116\ }
\begin{document}

\title{The enigmatic He-sdB pulsator \ls: \\
new insights from the VLT\thanks{Based on observations collected at the European Organisation for Astronomical Research in the Southern Hemisphere, Chile (proposal ID 093.D-0680 and 087.D-0950)}}
\author{
S.K. Randall \inst{1}
\and S. Bagnulo \inst{2}
\and E. Ziegerer \inst{3}
\and S. Geier \inst{1}
\and G. Fontaine \inst{4}
}

\institute{
ESO, Karl-Schwarzschild-Str. 2, 85748 Garching bei M\"unchen, Germany;
\email{srandall@eso.org} 
\and Armargh Observatory, College Hill, Armargh BT61 9DG, Northern Ireland, UK
\and Dr. Remeis-Observatory \& ECAP, Astronomical Institute, Friedrich-Alexander University Erlangen-N\"urnberg, Sternwartstr. 7, 96049 Bamberg, Germany
\and D\'epartement de Physique, Universit\'e de Montr\'eal, C.P. 6128,
Succ. Centre-Ville, Montr\'eal, QC H3C 3J7, Canada 
}
\date{Received date / Accepted date}

\abstract
{The intermediate Helium subdwarf B star \ls is a unique object showing extremely peculiar atmospheric abundances as well as long-period pulsations that cannot be explained in terms of the usual opacity mechanism. One hypothesis invoked was that a strong magnetic field may be responsible. We discredit this possibility on the basis of FORS2 spectro-polarimetry, which allows us to rule out a mean longitudinal magnetic field down to 300 G. 

Using the same data, we derive the atmospheric parameters for \ls to be $T_{\rm eff}$ = 35,150$\pm$111 K, $\log{g}$ = 5.88$\pm$0.02 and $\log{N(\rm He)/N(\rm H)}$ = $-$0.62$\pm$0.01. The high surface gravity in particular is at odds with the theory that \ls has not yet settled onto  the Helium Main Sequence, and that the pulsations are excited by an $\epsilon$ mechanism acting on the Helium-burning shells present after the main Helium flash.

Archival UVES spectroscopy reveals \ls to have a radial velocity of 149.1$\pm$2.1 km/s. Running a full kinematic analysis, we find that it is on a retrograde orbit around the Galactic centre, with a Galactic radial velocity component $U$=13.23$\pm$8.28 km/s and a Galactic rotational velocity component $V$=$-$55.56$\pm$22.13 km/s. This implies that \ls belongs to the halo population, an intriguing discovery. 
}

\titlerunning{The He-sdB pulsator \ls}
\authorrunning{S.K. Randall et al.}
\maketitle

\section{Introduction}    

Pulsating subdwarf B (sdB) stars have received a lot of attention over the last years. Having become one of the success stories of asteroseismology, they hold the key to a more mature understanding of the formation and evolution of hot subdwarfs. While it is commonly accepted that sdB stars are compact, evolved, He-core burning objects that lost too much of their H-envelope before or at the He-flash to sustain H-shell burning, the details of their evolution and in particular the mass loss mechanism remain unclear. A large fraction of field subdwarf B stars are found to reside in binary systems and may have been stripped of their envelope mass by binary interactions involving Roche lobe overflow and/or a common envelope phase \citep{han2002,han2003}, potentially resulting in a late Helium-flash \citep[e.g.,][]{brown2001}. Single sdB stars have been explained by mergers involving at least one white dwarf \citep{han2008,clausen2011} or a Helium-enriched subpopulation in Globular Clusters \citep[e.g.,][]{dantona2005}.

First discovered in the late nineties \citep{kilkenny1997}, the rapidly pulsating sdB stars show luminosity variations on timescales of 100-200 s and are found in a well-defined instability strip between $\sim$29,000-36,000 K. The fast pulsations have been modelled very successfully in terms of pressure ({\it p})-mode instabilities excited by an opacity ($\kappa$-)mechanism associated with a local overabundance of iron in the driving region \citep{charp1996,charp1997}. While sdB stars are overall metal-poor, diffusion (the competitive action between radiative levitation and gravitational settling) creates a non-uniform abundance profile of heavy elements as a function of depth. As a consequence, all sdB stars observed so far show chemically peculiar atmospheric abundances, and the vast majority of them ($\sim$90\%) are Helium-deficient (ranging from no detected Helium to nearly Solar).
Diffusion is also the key ingredient to understanding the pulsation driving in a well-defined instability strip in surface gravity - effective temperature ($\log{g}-T_{\rm eff}$) space: depending on the exact balance between the levitation of heavy elements caused by the radiation pressure (dependent on $T_{\rm eff}$) and the gravitational settling (dependent on $\log{g}$) an iron-"bump" is formed, causing a sharp peak in opacity at the right depth to drive pulsations, or not. The depth at which pulsations are excited depends critically on their type: the rapid $p$-mode pulsations are driven in the outer envelope regions, whereas in contrast the longer-period gravity modes ($g$-modes) are produced deeper inside the star. And indeed, the iron opacity mechanism also gives rise to a second type of sdB pulsator with longer ($P\sim$2000-8000 s) $g$-mode pulsations, members of which are located in a cooler instability strip between 22,000-29,000 K \citep{green2003}. It is then quite fitting that stars located at the intersection between the {\it p-} and {\it g-}mode instability regions appear to be hybrids and show both rapid and slow luminosity variations \citep{schuh2006}.  The two types of sdB pulsator have now been very successfully analyzed using asteroseismology \citep[e.g.,][]{val2013,charp2011}, where the internal parameters of the star are derived from the observed pulsation spectrum to a high accuracy and can be used to start constraining the proposed evolutionary scenarios. 

Given that the pulsation properties of the two types of sdB pulsator are so well defined, it was a surprise when long-period ($P\sim$ 2000-5000 s), multi-periodic luminosity variations were detected in the hot ($T_{\rm eff}\sim$35,000 K) sdB star LS IV$-$14$^\circ$116 \citep{ahmad2005,green2011}. While exhibiting pulsation periods typical of the $g$-mode pulsators, this unique star is located right in the middle of the hotter $p$-mode instability strip \citep[see Fig. 5 of][]{green2011}, where the longer pulsations observed cannot be explained by the $\kappa$-mechanism. As an alternative, it was suggested that the observed pulsations may be excited by an $\epsilon$-mechanism acting in the He-burning shells that appear before the star settles in the core He-burning stage \citep{miller2011}.  

 Compared to the other, H-rich sdB pulsators, \ls is also different in that it shows a mild atmospheric Helium-enhancement ($\log{N(\rm He)/N(\rm H)}\sim -$0.6). This makes it an intermediate He-rich sdB, and as such a potential asteroseismic probe of a different evolutionary phase to the other sdB pulsators. For instance, it has been suggested that He-sdBs are the immediate progenitors of H-sdBs, before the Helium has had time to be depleted from the atmosphere due to gravitational settling \citep{jeffery2012}. Apart from being the only He-rich sdB pulsator, \ls also shows a very strange abundance pattern even for a subdwarf B star, with Ge, Sr, Y and Zr enhanced by factors of up to 10,000 compared to Solar \citep{naslim2011}. One of the most plausible explanations invoked for these observed overabundances was radiative levitation in a particularly quiet atmosphere invoked by a strong ($\gtrsim$ kG) magnetic field. Such a magnetic field would presumably affect the driving of pulsation modes, and potentially shift the $g$-mode instability strip to higher temperatures, thus naturally explaining the pulsations observed in \ls.  

The main aim of the study presented here was to check LS IV$-$14$^\circ$116 for the presence of a magnetic field with a disk-averaged longitudinal component of a few hundred Gauss or higher using the spectro-polarimetric capability of FORS2 at the VLT on Cerro Paranal, Chile. This facility has over the last decade been used extensively to study stellar magnetism, and led to claims of magnetic field detections in a variety of stars across the Hertzsprung-Russell diagram, including hot subdwarfs \citep{otoole2005}. However, several of these detections (including those presented for sdB stars) were later shown to be artefacts of the data reduction procedure in a systematic analysis of FORS2 spectro-polarimetric archival data \citep{bagnulo2012,landstreet2012a}. An independent analysis based on spectro-polarimetry obtained with ESPaDOnS at the CFHT \citep{petit2012} also failed to measure a magnetic field in two hot subdwarfs for which a positive detection had previously been claimed. In this context, it was with great interest that we noted the first detection of a very strong (several hundred kG) magnetic field in a He-rich subdwarf O star \citep{heber2013}. Given that magnetic fields are relatively common among white dwarfs, they should also be present for a sizeable fraction of their hot subdwarf (sdB and sdO) progenitors. With its puzzling pulsational properties and atmospheric abundance profile, \ls was then a prime candidate for the first magnetic sdB star. 

In the following sections we present the FORS2 spectro-polarimetry gathered and the resulting null detection of a magnetic field. We then move on to the exploitation of the FORS2 total intensity spectra and supplement these with UVES archival data to show that \ls is very likely a halo star. The paper ends with a discussion of the open questions regarding this unique object. 

\section{No evidence for a magnetic field}

\begin{figure*}[t]
\centering
\includegraphics[width=14.0cm,angle=270]{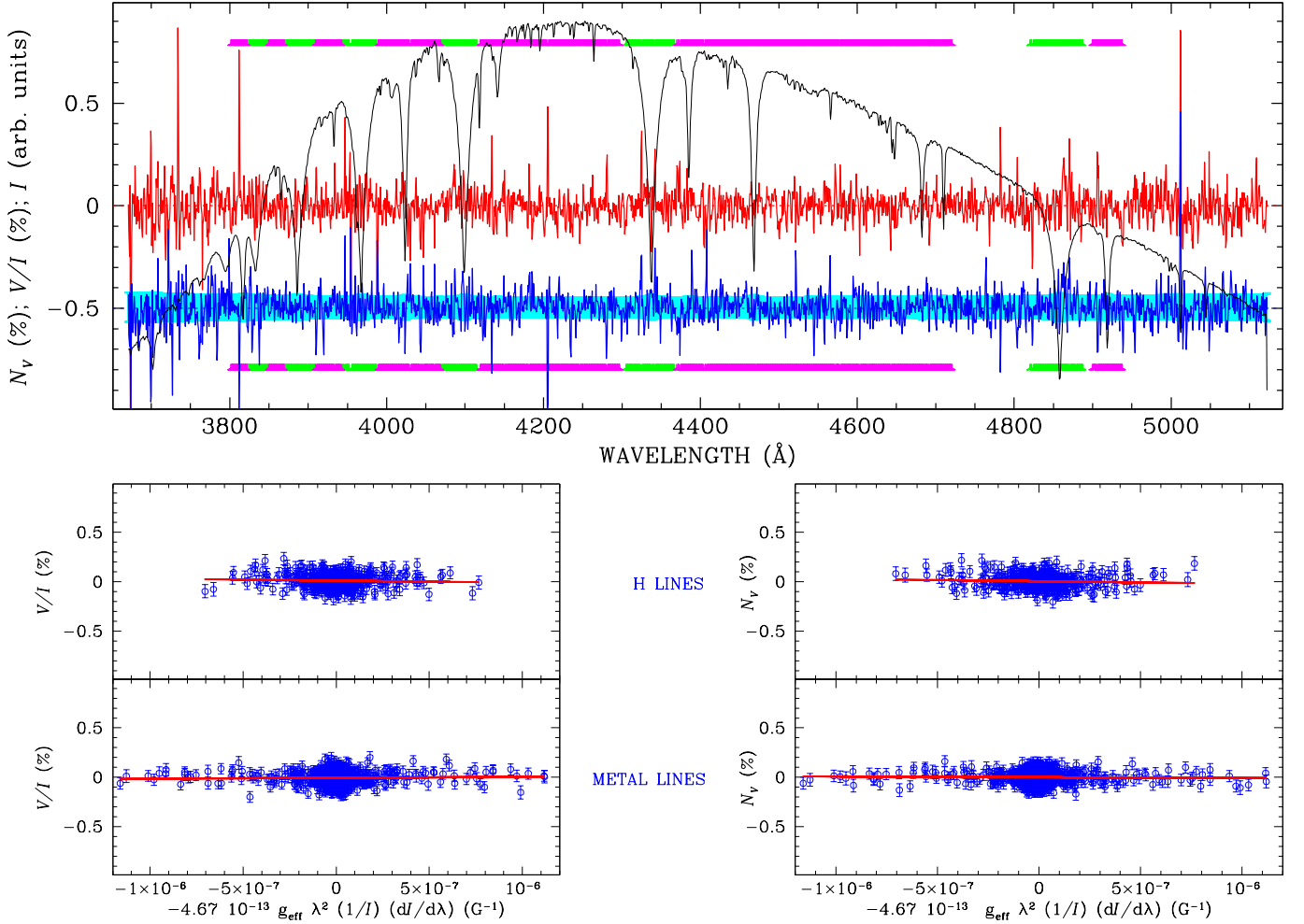}
\caption{{\it Top:} The total intensity FORS2 spectrum $I$ (black) together with the normalised circular polarisation $P_V$ (red) and the Null profile $N_V$ (blue) obtained by combining all data taken on 2014-06-05 for \ls. The null profile is offset by $-$0.5\% for display purposes, the 1$\sigma$ errors expected from photon noise being represented by the light blue error bars. The green (magenta) marked wavelength ranges indicate the portions of the spectrum used for the analysis based on the H (metal) lines respectively. {\it Bottom:} $P_V$ (left) and $N_V$ plotted as a function of the quantity indicated on the x-axis. The slope of $P_V$ in this diagram is proportional to the mean longitudinal magnetic field ${\langle\rm{B}_z\rangle}$. The plots refer to the results obtained from using the H lines (green portions of the wavelength range in the top plot) and the metal lines (magenta portions) as indicated.
}
\label{ls_pn}
\end{figure*}

Our spectro-polarimetric data were gathered with FORS2/VLT in the second half of the night on 4th and 5th June 2014 in visitor mode, which allowed us to use the blue-sensitive E2V CCD. We chose the 1200B grism covering the $\sim$3600-5200 \AA\ range and a variable slit width of 1$\arcsec$ for June 4 and 0.7$\arcsec$ for June 5, yielding a nominal wavelength resolution of $\Delta\lambda\sim$ 3.1 \AA\ and 2.3 \AA\ for the two nights respectively. The slit width was adjusted to limit slit losses due to the poor seeing (~1.5-2$\arcsec$) on the first night and kept at the originally desired value on the second night (seeing of 0.8-1.2$\arcsec$). We employed a shutter time of 600 s per exposure and set up the observing sequence so as to alternate between  two different retarder waveplate settings of $-$45$^\circ$ and +45$^\circ$. On the first night we obtained 16 exposures (8 at each of the retarder plate settings), while on the second night we managed to collect 30 spectra (15 pairs at perpendicular retarder plate settings). In total, we gathered nearly 8 hours of useful data on-target, the aim being to reach a S/N$\gtrsim$2000/\AA\ for this relatively faint ($B\sim$13.0) target. We also observed the bright spectropolarimetric standard HD 94660 using the exact same instrument setup as a sanity check.  

Data reduction was done closely following the procedure developed and described by \citet{bagnulo2012}. Simply put, we calculated two profiles: the circular polarisation normalised to the intensity $P_V$ = $V/I$ (where $V$ is the Stokes parameter measuring the circular polarization and $I$ is the usual unpolarized intensity), and the null profile $N_V$, which is representative of the noise of $P_V$. We obtained the 'rectified' version of these quantities by combining the series of exposure pairs taken at perpendicular retarder waveplate settings using Eq. (3) of \citet{bagnulo2012}. The results are illustrated in Fig. \ref{ls_pn} for the second night of observations (the result for the first night is qualitatively the same, but the S/N is lower). The top panel shows the total intensity (i.e. the time-averaged spectrum) together with $P_V$ and $N_V$ as a function of wavelength. Even more interesting are the diagnostic plots in the lower panels, as is explained in what follows. The maximum S/N in the spectrum is $\sim$2500 per \AA. 

Under the so called ``weak-field'' approximation, the intensity of circular polarisation (due to the Zeeman effect) across a spectral line can be expressed as:

\begin{equation}
P_V = -g_{\rm eff} \ C_z \ \lambda^{2}
\frac{1}{I} \frac{{\rm d}I}{{\rm d}\lambda} {\langle \rm{B}_z\rangle}
\end{equation}

\noindent where $g_{\rm eff}$ is the effective Land\'{e} factor, $\lambda$ is the wavelength, $C_z$ is a constant and ${\langle\rm{B}_z\rangle}$\ is the mean longitudinal magnetic field expressed in G. The longitudinal field is the integrated line-of-sight component of the magnetic field. As can be deduced from Eq. (1), the latter is obtained as the slope of a linear regression of $V/I$ versus the quantity $-g_{\rm eff} \ C_z \ \lambda^{2} \ (1/I) \ {\rm d}I/{\rm d}\lambda$, which is precisely what is plotted in the bottom left-hand panel of Fig.\ref{ls_pn}. Note that for the hydrogen lines of a star with $T_{\rm eff}\sim$35,000 K and a dipolar field, the ``weak-field'' approximation is formally valid for field strengths up to 10-20 kG \citep[see, e.g. Eq. 1 from][]{bagnulo1995}.

A visual inspection of $P_v$ yields no evidence for the presence of a strong magnetic field. Using the least-squares technique discussed in \citet{bagnulo2002} and applying it to all spectral lines we measure $\langle B_z\rangle$ = 9$\pm$93 G, a null detection within the error bar of about 100 G. From the H lines alone we derive $\langle B_z \rangle$ = $-$180$\pm$205 G and from the He and metal lines we find $\langle B_z \rangle$ = 97$\pm$105. The observations obtained during the first night also resulted in a null detection, although with a larger uncertainty due to the smaller S/N ratio of the spectra. As expected, null detections were also obtained from the null profiles, confirming the reliability of our results. However, in Fig.~1 both the $P_v$ and $N_v$ profiles show more scatter around zero than would be expected from photon noise alone (i.e. the scatter in the profiles is larger than the light blue 1-$\sigma$ error bars). A deeper inspection of our data reveals that spectra obtained during an observing series are slightly offset from each other. If interpreted in terms of radial velocity, the observed shifts would suggest that during the observing series, the star has changed its radial velocity by about 40 km/s, which is probably unrealistic and most likely due to instrumental effects. We note that during the observing series the instrument rotated by about $120^\circ$ while the airmass changed from about 1.5 to 1.1. In these conditions, tests performed in imaging mode show that a differential shift of about 1 pixel is to be expected (see Fig. 2.4 of the FORS user manual); the presence of a grism and polarimetric optics will presumably strengthen this effect. We conclude that the observed shift is probably due to to small instrument flexures, which in turns result in Stokes profiles noisier than expected from photon noise. For more detailed information see \citet{bagnulo2013}.

\begin{figure*}[t]
\centering
\begin{tabular}{c}
{\includegraphics[width=10.0cm,angle=270]{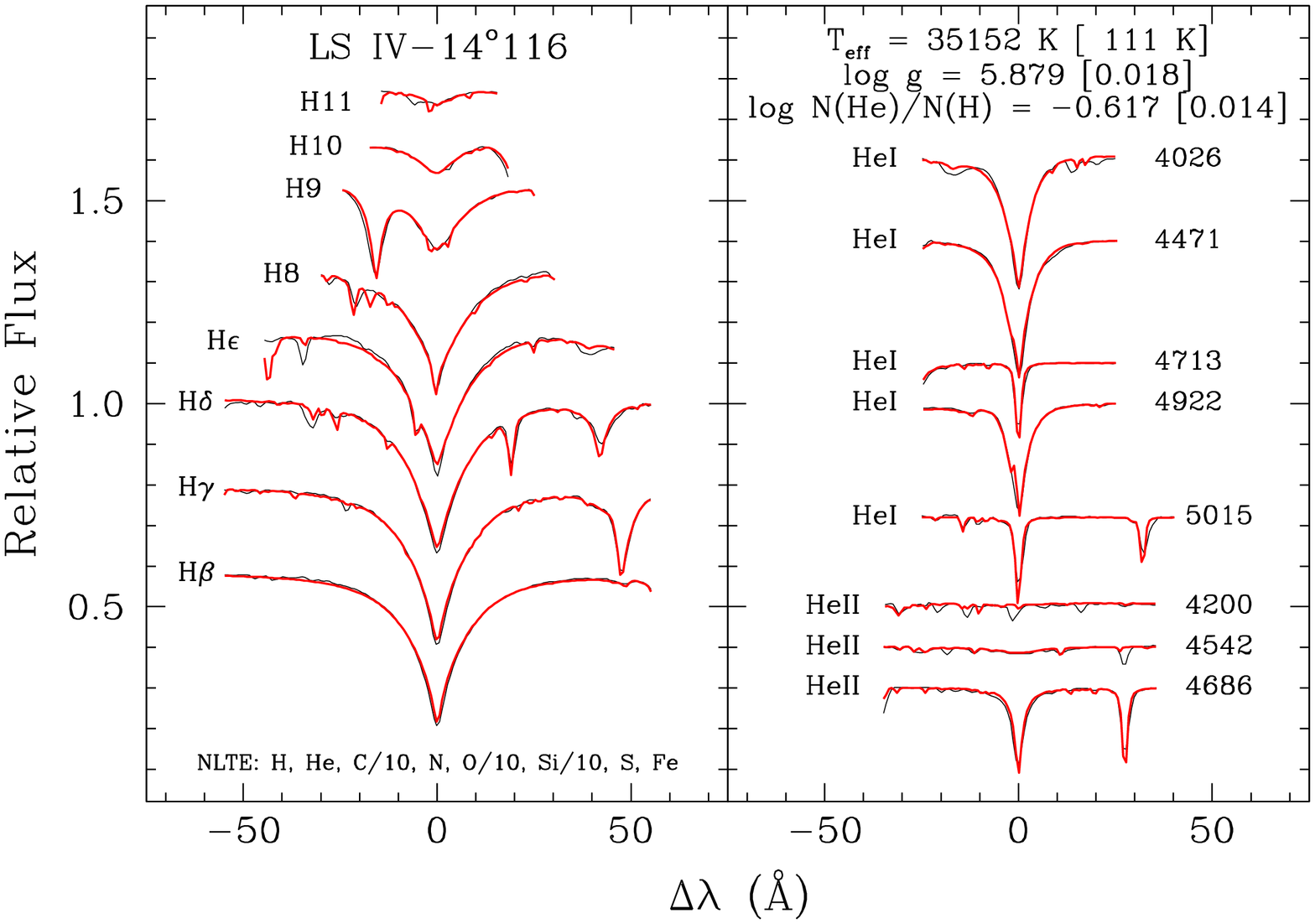}} \\ 
{\includegraphics[width=10.0cm,angle=270]{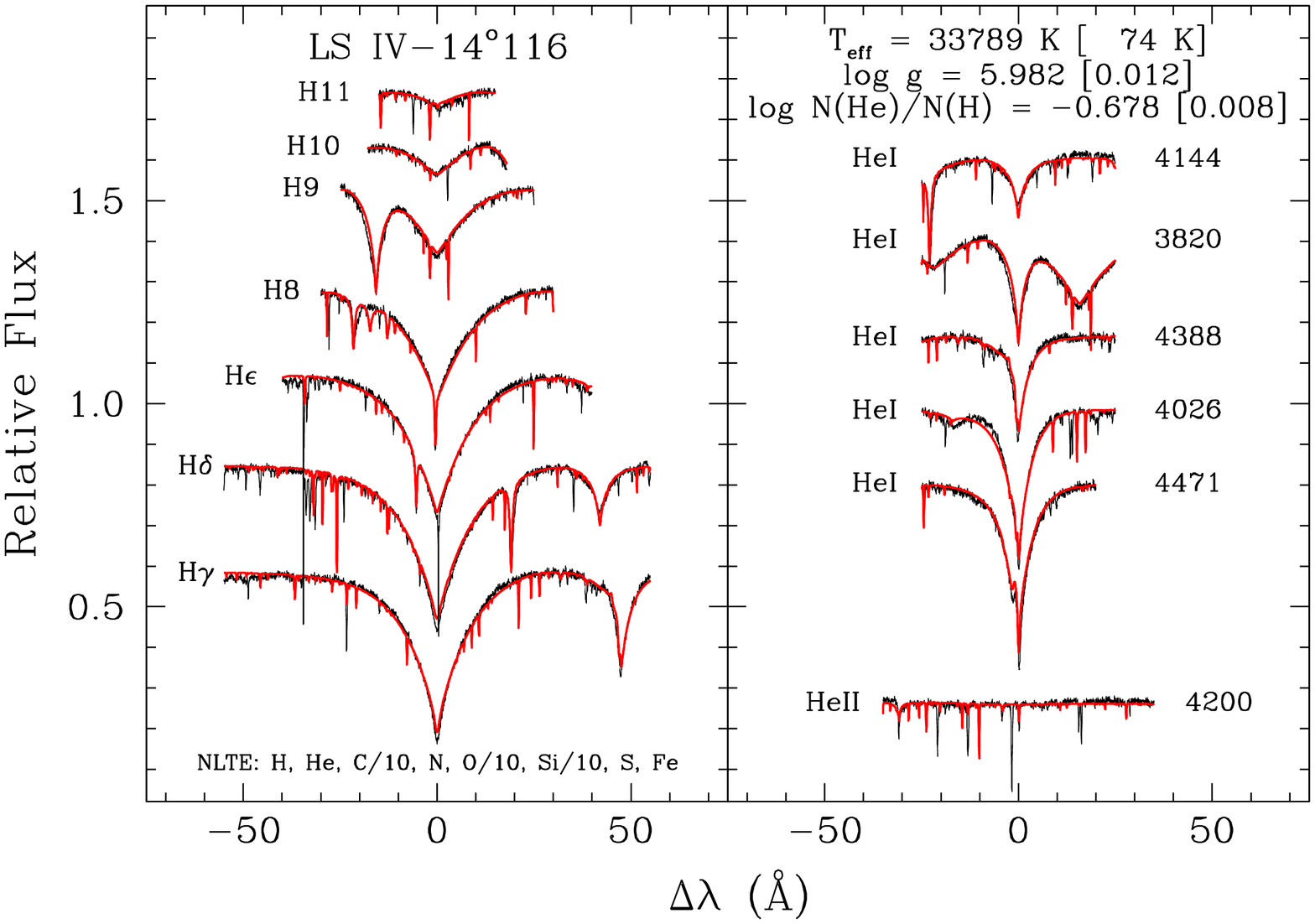}}
\end{tabular}
\caption{{\it Left:} Fit to the Balmer and He lines in the FORS2 spectrum of \ls. The thin black line is the observed spectrum and the  thick red line represents the best-fit model atmosphere. The quoted uncertainties refer only to the formal fitting errors, and certainly underestimate the true errors. {\it Right:} The equivalent fit to the UVES spectrum.}
\label{specs}
\end{figure*}   

From the analysis above we claim a null detection of a magnetic field down to a 3$\sigma$ threshold of 300 G. Note that, formally, this applies only to the mean longitudinal magnetic field (i.e. the magnetic field averaged over the visible disk in the direction of the line of sight) at the time of observation. The possibility remains that \ls in fact does have a strong dipolar field but that its longitudinal component was not detectable because we observed the magnetic field equator-on. Since a detailed analysis of the pulsation periods of \ls \citep{green2011} showed no indication of rotational splitting, we can assume that our target is a slow rotator with a rotation period longer than a month or so. Therefore, our spectropolarimetric measurements over two consecutive nights would have been taken at a very similar rotational phase. Assuming a limb darkening coefficient of 1, the mean longitudinal field of a dipole is related to the field strength at the magnetic pole $B_{\rm p}$ by
\begin{equation}
\langle \rm{B}_z\rangle = 0.4 B_p \cos{\ell}
\end{equation}
where $\ell$ is the angle of the dipolar axis with respect to the line of sight \citep[see Eqs. 2 and 3 of][]{bagnulo1995}. If we observed the magnetic field pole-on we would then (at 3 $\sigma$) have been sensitive to polar magnetic field strengths down to $\sim$ 750 G, while at $\ell$=45$^\circ$ we would have barely detected a field with $B_{\rm p}\sim$ 1 kG. A dipole magnetic field with $B_{\rm p}\gtrsim$ 5 kG would then have had a 85\% chance, one with $B_{\rm p}\gtrsim$ 7.5 kG a 90\% chance of being detected from our measurements.

Another issue is that the technique used is sensitive mainly to the longitudinal component of a dipolar field. According to Eq. 5 of \citet{landolf1998}, and again assuming a limb-darkening coefficient = 1, the contribution to the longitudinal field of any quadrupolar component is at best about six times smaller than the contribution due to the dipolar component. This implies that a quadrupolar field would have to be very strong indeed (at the 50 kG level or higher) to be detected by our measurements. According to the work of \citet{mathys2006} there is an empirical correlation between a dipole and a quadrupole magnetic field (which can be measured on the basis of line profiles) for roAp stars, i.e. stars with larger quadratic fields also have larger dipole fields. However, it is completely unclear whether such a correlation would hold for hot subdwarfs. The authors also point out that for a magnetic field to be detectable the organisation has to be large-scale enough for contributions of various parts of the stellar surface to not cancel out, and the geometry of the magnetic field along the line-of-sight has to be favourable.

The strong magnetic fields so far measured in white dwarfs tend to be dipole fields with strengths on the order of several MG, with weaker fields at the tens of kG level also having been detected in a small number of these stars \citep{landstreet2012b,kawka2012}. Similarly, the recently discovered magnetic He-sdO star \citep{heber2013} has a magnetic fields strong enough to induce clear Zeeman splitting (hundreds of kG). We would have been extremely unlucky to have missed anything remotely comparable to these in \ls.

Our measurements for \ls can be contrasted with those we obtained for HD 94600, a bright Ap star with a well-measured magnetic field \citep[e.g.,][]{landstreet2014}. Here, the exact same least-squares fitting technique yields an integrated line-of-sight magnetic field of $\langle B_z \rangle\sim$ $-$2200 G. This is in line with predictions for the time of our observations (MJD$\sim$56813) assuming the candidate rotational period of the star of 2800 d (see Fig. 5 of Landstreet et al. 2014). Thus, the instrument setup and technique yielding the null detection reported above for \ls appear sound.

\begin{figure}[t]
\centering
\includegraphics[width=9.0cm,angle=0]{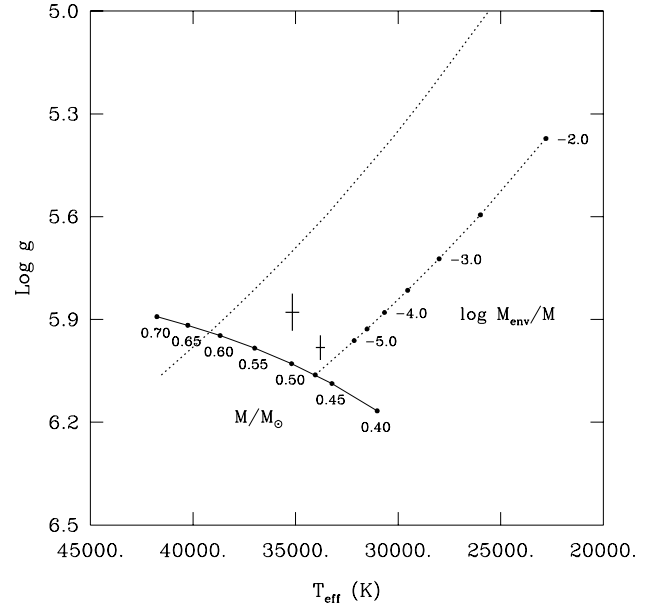}
\caption{We show the location of the Helium Main Sequence (solid line) in the $\log{g}-T_{\rm eff}$ diagram for different stellar masses as indicated. The Zero-Age EHB (right-hand dotted line) and the Terminal-Age EHB (left-hand dotted line) are shown for different hydrogen envelope masses based on the Montr\'eal 3rd generation evolutionary models. Superimposed on this is the location of \ls derived from the FORS2 (left) and UVES (right) spectra respectively. Both estimates place \ls firmly on the EHB; a star that has not yet settled on the EHB would be found at similar temperatures but significantly lower surface gravities \citep[cf. Fig 1,][]{miller2011}.}
\label{tracks}
\end{figure}

\section{Spectroscopic analysis}

While the main goal of the FORS2 observations proposed for \ls was the measurement of a possible magnetic field, the total intensity (Stokes $I$) spectra are also of interest. In particular, we planned to use the combined time-averaged spectrum obtained to determine the atmospheric parameters, and hoped to exploit the time-series of individual spectra for radial velocity (RV) measurements tracing the stellar pulsations. The latter study eventually proved fruitless due to the $\sim$40 km/s RV drift over the course of the nightly observations discussed above. Attempts to correct for the instrumental drift yielded RV curves too noisy to detect pulsations at the expected few km/s level. Quite interestingly, we did note a strong absolute RV shift of around $-$150 km/s, which was either overlooked or dismissed as unimportant in previous studies of \ls \citep{ahmad2005,green2011,naslim2011}. We were able to confirm this blueshift from a re-analysis of high resolution UVES spectra downloaded from the ESO archive. The UVES dataset comprises a 4-hour time-series of around 30 spectra that were originally obtained with the aim of detecting pulsational radial velocity variations, and the details of this are published elsewhere (Jeffery et al. 2014, MNRAS, submitted). For the purposes of the present study, we focus exclusively on the time-averaged absolute radial velocities and the combined spectrum.   

\subsection{Atmospheric parameters}

\begin{figure*}[t]
\centering
\includegraphics[width=12.0cm,angle=0]{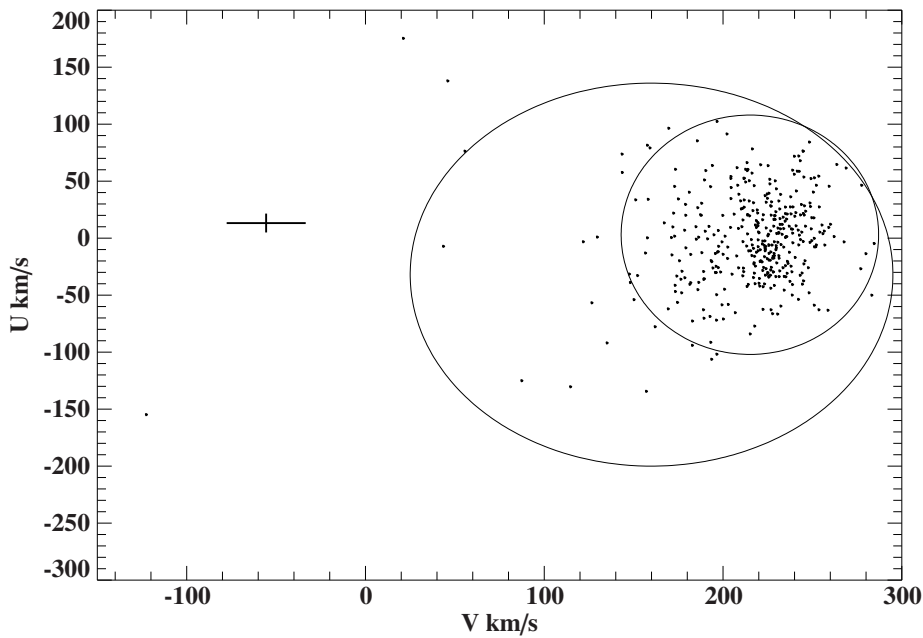}
\caption{U-V diagram showing the position of \ls (thick error bars) as compared to 3-$\sigma$ thin disk and 3-$\sigma$ thick disk contours. Stars located outside these contours have kinematic properties of the halo population. The points inside the contours correspond to the reference white dwarf sample of \citet{pauli2006}. These authors performed a detailed kinematic analysis of 398 DA white dwarfs using the same method as applied here.}
\label{uv}
\end{figure*}

The atmospheric parameters so far derived for \ls by two independent groups \citep{naslim2011,green2011} are somewhat at odds with each other, in particular as far as the surface gravity derived is concerned. On the basis of their high resolution echelle spectra, \citet{naslim2011} measured $T_{\rm eff}$=34,500$\pm$500 K and $\log{g}$=5.6$\pm$0.2, while \citet{green2011} report $T_{\rm eff}$=34,950$\pm$250 K and $\log{g}$=5.93$\pm$0.04. We made use of the combined spectrum from both our FORS total intensity spectra and the UVES archival data to re-evaluate the atmospheric parameters of \ls. 

The FORS2 combined spectrum was computed using our individual spectra obtained on June 5 only, i.e. those with the better spectral resolution of $\Delta\lambda\sim$ 2.3 \AA. This yielded a high quality time-averaged spectrum covering the 3650-5100 \AA\ range, including the Balmer lines from H$\beta$ upward as well as 8 strong Helium lines. These were fit using our standard grids of NLTE model atmospheres and synthetic spectra \citep[see e.g.][for details]{brassard2010} incorporating a fixed metallicity inspired by the abundance analysis of \citet{blanchette2008}. Note that these are the same models as used by \citet{green2011} in their analysis, and that the metal abundances used are representative of hot subdwarfs in general, but were not tailored to the extremely peculiar abundance profile of \ls. The combined UVES spectrum was analysed using the same models, but this time convolved to the appropriate 0.1 \AA\ wavelength resolution. Here, we combined the blue part of the pipeline-reduced individual spectra as available from the ESO Phase 3 data products archive.   

The best-fit synthetic spectra are shown together with the observed FORS2 and UVES spectra and the inferred atmospheric parameters in Fig. \ref{specs}. While the values inferred are not compatible within the formal errors (which are well known to underestimate the true uncertainties), they both yield a value of $\log{g}$ close to that inferred by \citet{green2011}. Given the errors almost certainly introduced when combining the echelle orders and fixing the continuum of the UVES spectra, we believe the FORS2 values to be more reliable. Therefore, we adopt $T_{\rm eff}$=35,150 K and $\log{g}$=5.88 in what follows. These values place \ls firmly on the He-burning main sequence, as can be seen in Fig. \ref{tracks}.

\subsection{Kinematics}

We searched the series of archival UVES spectra for RV variations by measuring the Doppler shift of the spectral lines with respect to their rest wavelength. This was achieved by fitting a set of mathematical functions with the FITSB2 routine \citep{napi2004} to the He I lines at 4026.2, 4120.9 and 4387.9 \AA. We used a polynomial to match the continuum, while the line wings were fit with a Lorentz profile and the line core with a Gaussian profile.  Heliocentric corrections were then applied to the times and velocities.

Averaging the RV measurements from the individual UVES spectra we infer an absolute average radial velocity shift of $v_r=-149.1\pm$2.1 km/s. This is in line with the $\sim -$150 km/s shift seen in our FORS2 spectra. Note also that the UVES spectra do not show a radial velocity drift comparable to that found for the FORS data, confirming the instrumental nature of that effect.  Upon reinspection, the lower spectral resolution ($\Delta\lambda =$ 9 \AA) Bok telescope spectra used for the spectral analysis of \citet{green2011} show a compatible wavelength shift. There can then be no doubt that the blueshift detected is real. Since single subdwarf B stars in the Galactic field tend to have significantly lower radial velocities ($|RV|<$ 40 km/s, Betsy Green, 2014, private communication) we found this highly intriguing and decided to run a kinematic analysis.

The velocity components of \ls were computed based on its coordinates ($\alpha_{2000}$ = 20h57m38.8s, $\delta_{2000}$ = $-$14$^{\circ}$25$\arcmin$47$\arcsec$ from Simbad), radial velocity (determined from the UVES spectra above), proper motion and distance. All the parameters were varied within the errors adopted.  The distance was estimated to be $d=$ 0.44$\pm$0.04 kpc following the method described by \citet{ramspeck2001} and assuming the values of $\log{g}$ and $T_{\rm eff}$ as inferred from our FORS2 spectra shown in the left panel of Fig. \ref{specs} (but increasing the errors to more realistic values of 0.05 and 500 K respectively), a canonical mass of 0.47 $M_{\odot}$, a magnitude of $V=$ 13.03 from Simbad, and using the extinction maps of \citet{schlafly2011}. The proper motion was taken from the PPMXL catalogue via Vizier to be $\mu_{\alpha}\cos\alpha$ = 8.2$\pm$1.8 mas/yr and $\mu_{\delta}$ = $-$130.6$\pm$1.8 mas/yr.  Since these observed parameters are in the heliocentric reference frame, they need to be supplemented by Solar velocity and position information in order to be transferred to the Galactic rest frame. Here we assumed the distance of the Sun from the Galactic centre to be 8.4 kpc, its motion relative to the Local Standard of Rest (LSR) to have components $v_x\sun$=11.1 km/s, $v_y\sun$=12.24 km/s, $v_z\sun$=7.25 km/s \citep{schoenrich2010}, and the velocity of the LSR to be $V_{lsr}$ = 242 km/s, as predicted by Model I of \citet{irrgang2013}. 

The position and kinematic properties of \ls were then computed following the equations given in Appendix A. We derived Galactic coordinates $x$ = $-$8.10$\pm$0.03 {\bf kpc}, $y$ = 0.20$\pm$0.02 kpc, $z$ = $-$0.25$\pm$0.02 kpc and a Galactic restframe velocity of 60.39$\pm$19.37 km/s, with cartesian velocity components $v_x = -$14.66$\pm$7.65 km/s, $v_y = -$55.23$\pm$22.28 km/s and $v_z$=$-$10.21$\pm$9.84 km/s. This translates to a Galactic radial velocity component $U$=13.23$\pm$8.28 km/s and a Galactic rotational velocity component $V = -$55.56$\pm$22.13 km/s. The negative value of $V$ immediately indicates a retrograde orbit, i.e. \ls is orbiting the Galactic centre in the opposite direction to the disk, implying it to be a halo star. Fig. \ref{uv} illustrates this quite nicely: while the stars of the reference sample fall mostly within the 3$\sigma$ contours expected for the thin or thick disk population in the $U-V$ plane, \ls lies well outside these contours. It must therefore be a halo star that by chance is relatively close by. For completeness, it should be noted that using the atmospheric parameters and in particular the lower surface gravity of \citet{naslim2011} for the distance estimate gives an even lower value of $V\sim -$150 km/s, i.e. the qualitative conclusion that \ls is a halo star does not change.

\section{Conclusion}

The conclusions we can draw from the study presented here are: 
\begin{itemize}
\item \ls does not have a strong magnetic field. Formally, the FORS2 spectro-polarimetric measurements exclude a mean longitudinal magnetic field higher than 300 G.
\item The atmospheric parameters derived from the combined FORS2 total intensity spectrum are consistent with \ls residing on the Helium main sequence.
\item Based on a kinematic analysis, \ls appears to belong to the halo population of our Galaxy.
\end{itemize}

The fact that \ls does not have a discernable magnetic field means that we need to find alternative scenarios to explain the pulsations as well as the extremely peculiar abundance pattern. One currently popular idea is that \ls is a pre-Helium Main Sequence (pre-He-MS) star that is still in the process of settling onto the Extreme Horizontal Branch. In this scenario, the observed abundance pattern can be qualitatively understood in terms of an atmosphere not yet sorted by diffusion. Diffusion, if unimpeded by other processes, will drain the atmosphere of a hot subdwarf of metals (and even Helium if any Hydrogen is present) on a time-scale of $\lesssim$10$^4$ years, negligible compared to the 10$^8$-year time-scale of the He-core-burning phase \citep[see, e.g. Fig. 11 of][]{latour2014}. It is clear that in order for He-rich sdBs to exist in the numbers observed, diffusion must be slowed down, possibly by internal turbulence or a weak stellar wind, however the details are far from being understood. 

The pre-He-MS scenario is supported by the qualitative modelling of the pulsations observed in \ls in terms of an $\epsilon$-mechanism acting in the He-burning shells that appear before the star settles in the core He-burning phase \citep{miller2011}. Depending on the exact model parameters used, periods overlapping those observed in \ls can be excited for models in the right temperature range around 35,000 K. However, maximum instability occurs at $\log{g}\sim$5.2-5.7 and pulsations cease for models with $\log{g}\gtrsim$ 5.8 since the $\epsilon$-mechanism is active only during the He-burning subflashes following the primary He-core flash, which stop once the star settles on the He-MS. According to the atmospheric parameters derived above, \ls resides on the He-MS with $\log{g}\sim$5.9.

There are then two options: either \ls has already settled on the He-MS, or our atmospheric parameters are wrong. While we have convincing reasons to believe our atmospheric parameters to be rather accurate for H-sdB stars in the 29,000-36,000 K range thanks to frequent cross-checks with asteroseismic modelling, it is possible that the atmospheric parameters inferred for \ls will change slightly when including metals appropriate for the peculiar abundance profile observed in the model atmospheres. Note however that the assumed metal abundance affects mostly the derived value of $T_{\rm eff}$ and has little impact on $\log{g}$. A more precise abundance analysis of \ls is currently planned by another group on the basis of UV-spectroscopy scheduled with HST, and will hopefully provide new insights on this.  

If \ls is confirmed as a bona fide helium core-burning star, the driving of the pulsations and the unusual surface composition remain an enigma. We believe that the halo membership may be of importance in this context. Indeed, the atmospheric properties of halo and globular cluster hot subdwarfs show systematic differences compared to those in the Galactic disk in terms of their relative distribution in $\log{g}-T_{\rm eff}$ space \citep[e.g.][]{latour2014,nemeth2012}. Moreover, there is first observational evidence that intermediate He-rich stars in general belong to the halo (P. N\'emeth, 2014, private communication). The pulsational properties of hot subdwarfs in some globular clusters and the field also appear to be different. For instance, the rapid sdB pulsators of the field population have not yet been discovered in a globular cluster despite systematic searches \citep[e.g.][]{randall2011}, and conversely, the fast sdO pulsators recently found in $\omega$ Cen do not appear to have counterparts among field sdO stars \citep{johnson2014}. It is not at all clear why this is the case, but must reflect differences in the internal structure of the stars in question. 

Presumably, the metallicity of the subdwarf progenitor population plays an important role. Apart from being metal-poor, the red giant progenitors of sdB stars in the halo are old compared to those in the Galactic disk and have a relatively low mass of around 0.8 $M_{\odot}$. The progenitor mass is in turn predicted to influence the core mass of the sdB \citep{han2002}. It was also suggested that different formation channels for sdB stars dominate depending on the age of the population, the white dwarf merger scenario becoming more important for older populations \citep{han2008}. Besides updated pulsation models including variable envelope compositions, a better observational characterisation of the different hot subdwarf populations is needed to address these issues. It would be particularly interesting to run kinematic analyses for all well-studied subdwarfs with unusually large absolute radial velocities and/or proper motions to assess whether they are in fact halo or Galactic disk stars. Statistically significant samples for these populations could then be subjected to detailed atmospheric analyses and contrasted with each other. Ultimately, the asteroseismic exploitation of halo sdB pulsators may be used to compare the masses and structural compositions of these stars with those of their Galactic disk counterparts.

\begin{acknowledgements}
S.K.R. gratefully acknowledges the help of the Paranal Science Operations, Engineering and Software staff, in particular Linda Schmidtobreick, with the observations. The program used for the kinematic calculations was kindly provided by Andreas Irrgang. We thank Betsy Green for sharing her unpublished radial velocities of hot subdwarfs. Further thanks go to Peter N\'emeth, Marcelo Miller Bertolami, Uli Heber and Marilyn Latour for interesting and insightful discussions, and Pascal Petit, St\'ephane Charpinet, and Val\'erie Van Grootel for their interest and motivation. We also acknowledge the PI of the archival UVES data, Simon Jeffery. This research has made use of the SIMBAD database, operated at CDS, Strasbourg, France.
\end{acknowledgements}

\bibliographystyle{aa}
\bibliography{ms}

\appendix
\section{Kinematic calculations}

We calculated the kinematic properties of \ls based on the input parameters given in section 3.2 using the equations detailed below. To begin with, the polar coordinates of the target are converted to cartesian coordinates using
\begin{align*}
C=d\cdot\left(\begin{array}{c}\cos{\alpha}\cos{\delta}\\\sin{\alpha}\cos{\delta}\\\sin{\delta}\end{array}\right)
\end{align*}

\noindent where $d$ is the distance of the star (from the Sun) and $\alpha$ and $\delta$ are the right ascension and declination respectively. This cartesian coordinate system is then rotated and shifted to $x,y,z$ coordinates in the Galactic coordinate system, defined by the Galactic centre being at (0,0,0), the Sun being at $x=-$8.4 kpc (i.e. its assumed distance from the Galactic centre), and the positive z-axis pointing towards the Galactic North pole, using
\begin{align*}
\left(\begin{array}{c}x\\y\\z\end{array}\right)=M\cdot C+\left(\begin{array}{c}-d_{\sun-GC}\\0\\0\end{array}\right)
\end{align*}

\noindent where $d_{\sun-GC}$ is the distance from the Sun to the Galactic centre and
\begin{align*}
M=\left(\begin{array}{l}V^T_\text{GC}\\V^T_y\\V^T_\text{NGP}\end{array}\right)
\end{align*}
\begin{align*}
V_\text{\tiny GC}=\left(\begin{array}{c}\cos{\alpha_\text{GC}}\cos{\delta_\text{GC}}\\\sin{\alpha_\text{GC}}\cos{\delta_\text{GC}}\\\sin{\delta_\text{GC}}\end{array}\right)
\end{align*}
\begin{align*}
V_\text{NGP}=\left(\begin{array}{c}\cos{\alpha_\text{NGP}}\cos{\delta_\text{NGP}}\\\sin{\alpha_\text{NGP}}\cos{\delta_\text{NGP}}\\\sin{\delta_\text{NGP}}\end{array}\right)
\end{align*}
\begin{align*}
V_y=-(V_\text{GC}\times V_\text{NGP})
\end{align*}

\noindent are the rotation matrix transformations applied to the $x,z,y$ axes respectively \citep{johnson1987}. We performed these transformations assuming Galactic Centre coordinates $\alpha_\text{GC}$ = 17:45:37.224, $\delta_\text{GC}$ = $-$28:56:10.23 and Galactic North Pole coordinates $\alpha_\text{GNP}$ = 12:51:26.282, $\delta_\text{GNP}$ = 27:07:42.01 as determined by \citet{reid2004}.

For the computation of the velocities, the individual observed velocity components are converted to cartesian coordinates separately and subsequently added to give the total velocity $v_C$ in the cartesian system:
\begin{align*}
v_{rc}=v_r\cdot\left(\begin{array}{c}\cos{\alpha}\cos{\delta}\\\sin{\alpha}\cos{\delta}\\\sin{\delta}\end{array}\right)
\end{align*}
\begin{align*}
v_\delta=\mu_\delta\cdot d\cdot\left(\begin{array}{c}-\cos{\alpha}\sin{\delta}\\-\sin{\alpha}\sin{\delta}\\\cos{\delta}\end{array}\right)
\end{align*}
\begin{align*}
v_\alpha=\mu_\alpha\cos\delta\cdot d\cdot\left(\begin{array}{c}-\sin{\alpha}\\\cos{\alpha}\\0\end{array}\right)
\end{align*}
\begin{align*}
v_C=v_{rc}+v_\delta+v_\alpha
\end{align*}
\noindent where $v_r$ is the measured radial velocity, $\mu_{\delta}$ is the observed proper motion in declination, and $\mu_{\alpha}\cos\delta$ is the observed proper motion in right ascension. The resulting cartesian velocity components are then transformed into velocity components $v_x,v_y,v_z$ in the Galactic coordinate system with
\begin{align*}
\left(\begin{array}{c}v_x\\v_y\\v_z\end{array}\right)=M\cdot v_C+\left(\begin{array}{c}v_{x \sun}\\v_{y \sun} + v_{lsr}\\v_{z \sun}\end{array}\right)
\end{align*}
\noindent where $v_{lsr}$ is the velocity of the Local Standard of Rest and $v_{x\sun},v_{y\sun},v_{z\sun}$ are the velocity components of the Sun in the Galactic reference frame. The total Galactic restframe velocity is then given by
\begin{align*}
v_\text{grf} &= \sqrt{{v_x}^2 + {v_y}^2 + {v_z}^2}
\end{align*}
\noindent The Galactic radial velocity and rotational velocity components of the star are then respectively computed as
\begin{align*}
U &= \frac{x v_x + y v_y}{\sqrt{x^2 +y^2}} \\
V &= -\frac{x v_y-y v_x}{\sqrt{x^2 +y^2}}
\end{align*}

\end{document}